\documentclass[]{mosi}

\usepackage{helvet}

\usepackage{amsmath} 
\usepackage{natbib}
\usepackage{graphicx}
\usepackage{subcaption} 

\usepackage[toc,page,header]{appendix}
\usepackage[utf8]{inputenc} 
\usepackage[T1]{fontenc}    
\usepackage{hyperref}       
\usepackage{url}            
\usepackage{booktabs}       

\usepackage{amsfonts}       
\usepackage{nicefrac}       
\usepackage{microtype}      
\usepackage{wrapfig}
\usepackage{amssymb}        

\usepackage{titletoc}
\usepackage{minitoc}

\usepackage{array}
\usepackage{etoolbox}

\definecolor{lightblue}{RGB}{200, 230, 255}  
\definecolor{headerblue}{RGB}{150, 200, 255} 

\usepackage{pgfplots}
\usepackage{xcolor}
\usepackage{float}
\usepackage{comment}
\usepackage{multirow}
\usepackage{makecell}
\usepackage{siunitx}
\usepackage{tikz}
\usepackage{pgf-pie}
\usepackage[export]{adjustbox}

\usepackage{ragged2e}
\usepackage{tabularx}
\usepackage{caption}
\usepackage{enumitem}
\usepackage{pifont}
\usepackage[hang,flushmargin]{footmisc}

\usepackage{tcolorbox}
\tcbuselibrary{breakable}
\tcbuselibrary{skins}

\usepackage{listings}

\usepackage{threeparttable}
\usepackage{setspace}

\usepackage[table]{xcolor}

\definecolor{oursgray}{gray}{0.95}

\definecolor{MossCyan}{HTML}{82D9FF} 
\definecolor{MossBlue}{HTML}{82B1FF}

\definecolor{tickG}{HTML}{00C853}
\definecolor{crossR}{HTML}{FF1744}

\newcommand{\faHome}{\raisebox{-0.2ex}{\includegraphics[height=2.0ex]{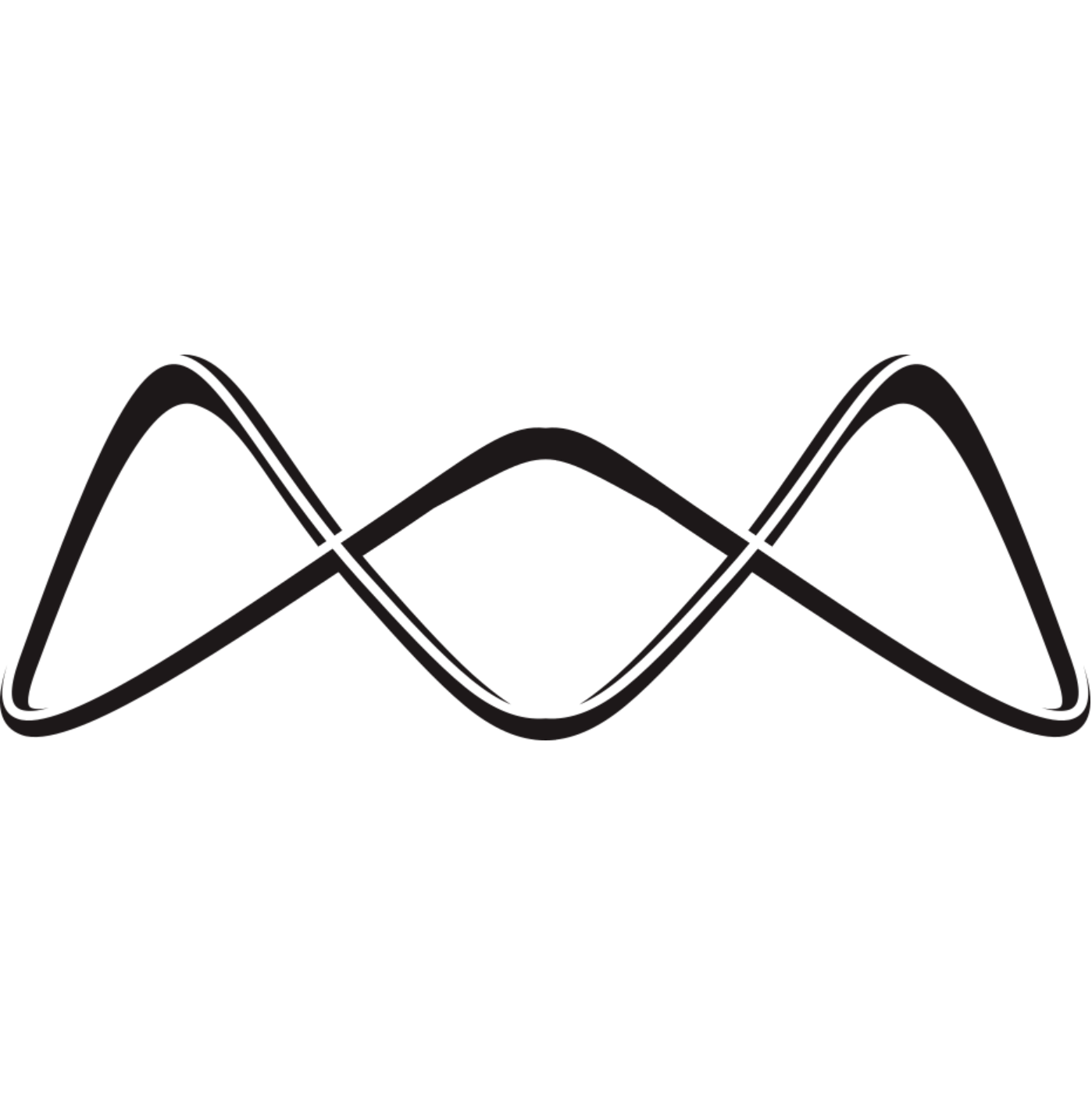}}}
\newcommand{\faPlayCircle}{\raisebox{-0.2ex}{\includegraphics[height=2.0ex]{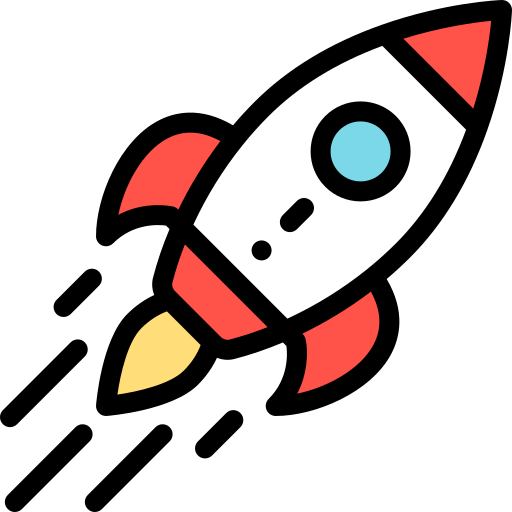}}}

\newcommand{\hflogo}{\raisebox{-0.2ex}{\includegraphics[height=2.0ex]{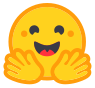}}}
\newcommand{\faGithub}{\raisebox{-0.2ex}{\includegraphics[height=2.0ex]{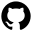}}}

\newtcolorbox{promptbox}[2][]{
    colback=white,
    coltext=black,
    arc=3mm,
    boxrule=0.5pt,
    colframe=black!60!white,
    title={#2},
    colbacktitle=black,
    coltitle=white,
    fonttitle=\bfseries,
    top=8pt,
    bottom=8pt,
    left=10pt,
    right=10pt,
    breakable,
    before upper={%
        \linespread{1}\selectfont
        \setlength{\parskip}{1ex plus 0.2ex minus 0.2ex}%
        \setlength{\parindent}{0pt}%
    },
    #1
}

\title{MOSS-VoiceGenerator: Create Realistic Voices with Natural Language Descriptions}

\author{SII-OpenMOSS Team\textsuperscript{*}}

\abstract{
Voice design from natural language aims to generate speaker timbres directly from free-form textual descriptions, allowing users to create voices tailored to specific roles, personalities, and emotions. Such controllable voice creation benefits a wide range of downstream applications—including storytelling, game dubbing, role-play agents, and conversational assistants, making it a significant task for modern Text-to-Speech models.
However, existing models are largely trained on carefully recorded studio data, which produces speech that is clean and well-articulated, yet lacks the lived-in qualities of real human voices.
To address these limitations, we present \textbf{MOSS-VoiceGenerator}, an open-source instruction-driven voice generation model that creates new timbres directly from natural language prompts. Motivated by the hypothesis that exposure to real-world acoustic variation produces more perceptually natural voices, we train on large-scale expressive speech data sourced from cinematic content. 
Subjective preference studies demonstrate its superiority in overall performance, instruction-following, and naturalness compared to other voice design models.
}

\checkdata[\raisebox{-0.1ex}{\faHome}\hspace{0.4em}Homepage]{\url{https://mosi.cn/models/moss-voice-generator}}
\checkdata[\raisebox{-0.1ex}{\faPlayCircle}\hspace{0.4em}Online Demo]{\url{https://huggingface.co/spaces/OpenMOSS-Team/MOSS-VoiceGenerator}}
\checkdata[\raisebox{-0.1ex}{\hflogo}\hspace{0.4em}Hugging Face]{\url{https://huggingface.co/OpenMOSS-Team/MOSS-VoiceGenerator}}
\checkdata[\raisebox{-0.1ex}{\faGithub}\hspace{0.4em}GitHub]{\url{https://huggingface.co/collections/OpenMOSS-Team/moss-tts}}

\begin{document}
\maketitle
\begingroup
\renewcommand{\thefootnote}{\fnsymbol{footnote}}
\setcounter{footnote}{1}
\footnotetext{Full contributors can be found in the Contributors section.}
\endgroup


\section{Introduction}

\noindent
In recent years, Text-to-Speech (TTS) systems have progressed from merely producing natural and high-fidelity speech toward supporting controllable and expressive speech generation~\citep{wang2023neural,anastassiou2024seed,zhou2025indextts2,du2024cosyvoice}. 
These advances enable a shift from traditional TTS or voice-cloning~(VC) to natural language–based instruction following task. 
By allowing users to control vocal characteristics, such as emotion, speaking style, and character persona, through free-form text descriptions, this paradigm substantially lowers the barrier to voice customization for non-expert users, which broadens the applicability of TTS to downstream scenarios such as audiobooks, game dubbing, role-play agents, and conversational assistants.

A growing body of work has explored instruction-driven voice generation, yet current approaches differ considerably in their degree of controllability, generalizability, and dependence on reference audio.
Early efforts in this direction still rely on reference speakers to anchor the generated timbre.
\citet{yang2024instructtts} introduces natural language style prompts for expressive TTS, but the model is still trained on a closed set of speakers, limiting generalization and preventing open-set timbre design.
Audiobox~\citep{vyas2023audiobox} leverages pretrained CLAP~\citep{wu2023large} embeddings to condition audio generation on text descriptions; however, CLAP models are typically trained on coarse-grained descriptions and struggle to capture fine-grained timbre nuances.
VoxInstruct~\citep{zhou2024voxinstruct} and EmoVoice~\citep{yang2025emovoice} leverage LLM understanding by concatenating style descriptions with synthesis text for fine-grained control, but still require reference speaker audio rather than generating timbres from scratch.
Moving toward fully reference-free generation, VoiceSculptor~\citep{hu2026voicesculptor} takes a step further by integrating instruction-based voice design with high-fidelity voice cloning in a unified framework, enabling controllable timbre generation from natural language descriptions with iterative refinement via Retrieval-Augmented Generation (RAG).
MIMO-Audio~\citep{zhang2025mimo} enhances controllability through massive pretraining followed by instruction tuning, demonstrating the promise of the LLM-based paradigm for open-domain voice design. The Qwen3-TTS family~\citep{Qwen3-TTS} also introduces models focused on voice design, further advancing the field and offering additional options for expressive and customizable speech synthesis.
On the commercial side, several APIs—including Elevenlabs, MiniMax, GPT-4o-TTS, and Gemini—have begun offering voice design or editing functionalities, reflecting growing market demand for instruction-driven timbre generation and customizable voice synthesis.

Despite significant recent progress in controllable speech synthesis, many existing models rely on studio-recorded high-quality audio, resulting in voices that sound clean and polished but lack the "lived-in" realism of everyday speech. The subtle imperfections of real speech, such as natural breath patterns, rhythmic irregularity, and spontaneous emotional coloring, remain largely absent, leaving a perceptible gap between synthesized voices and authentic human expression.
To address this, we introduce MOSS-VoiceGenerator, a fully open-source instruction-driven TTS model that generates realistic and expressive speech directly from natural language descriptions, without requiring any reference audio. By leveraging large-scale in-the-wild cinematic data and the strong instruction-following capabilities of decoder-only language models, MOSS-VoiceGenerator produces voices with greater naturalness and diversity.

Our main contributions are as follows:
\begin{itemize}
    \item We present \textbf{MOSS-VoiceGenerator}, a fully open-source instruction-driven TTS model that generates realistic and expressive speech directly from natural language descriptions, without requiring any reference audio.
    \item We construct a data pipeline suitable for instruction-based tasks, along with a richly annotated dataset sourced from cinematic content. Our developed embedding model and annotation model can effectively enable cost-effective data augmentation.
    \item Extensive evaluations show that MOSS-VoiceGenerator performs exceptionally well in subjective evaluations and achieves competitive performance in objective evaluations.
\end{itemize}

\section{MOSS-VoiceGenerator}

\subsection{Model Architecture}

\begin{figure}[t]
    \centering
    \includegraphics[width=0.85\textwidth]{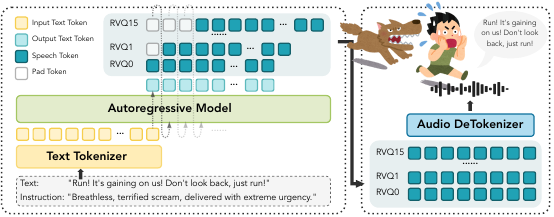}
    \caption{Illustration of the MOSS-VoiceGenerator inference. The voice description and text are concatenated and fed into a causal language model with delay-pattern generation; the output audio tokens are decoded by MOSS-Audio-Tokenizer.}
    \label{fig:model_structure}
\end{figure}

As illustrated in Fig.~\ref{fig:model_structure}, the architecture of MOSS-VoiceGenerator follows the design of MOSS-TTS~\citep{gong2026mossttstechnicalreport} with the delay pattern and utilizes the MOSS-Audio-Tokenizer~\citep{gong2026mossaudiotokenizerscalingaudiotokenizers} for audio tokenization. In our implementation, we only use the first 16 layers of the RVQ codebooks for training. The voice description and synthesis text are concatenated and input into the language model (LM). During inference, a time-delay shift is applied, where each codebook layer's tokens are shifted forward by $j-1$ frames relative to the first layer. After generating output until the end-of-sequence (EOS) token, the audio tokens are decoded into the final waveform.

This discrete autoregressive architecture offers several key benefits for the task. It unifies speech generation with language modeling as a sequence prediction task, allowing for the direct reuse of LLM training paradigms and inference frameworks. Autoregressive modeling naturally captures long-range dependencies, ensuring global prosody and timbre consistency. Additionally, the discrete framework eliminates the need for iterative inference, unlike diffusion or flow-matching models, leading to simpler and more efficient deployment. Finally, the shared sequence space for both speech and text tokens makes joint instruction-speech modeling more seamless, enhancing the model's instruction-following capabilities.

\subsection{Data Collection}

\begin{figure}[t]
    \centering
    \includegraphics[width=0.9\textwidth]{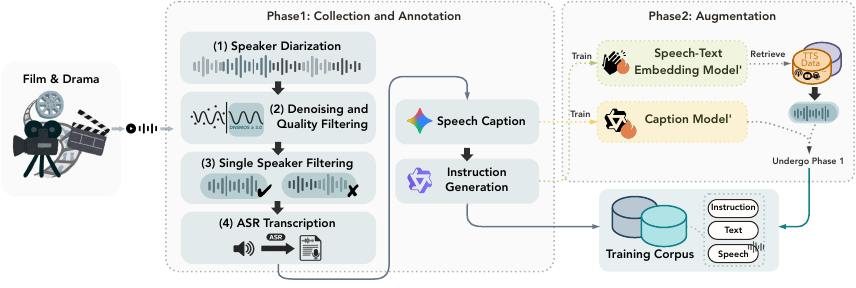}
    \caption{Data collection pipeline for MOSS-VoiceGenerator. Phase 1 annotates cinematic audio via speaker diarization, denoising and
quality filtering, single-speaker filtering, and ASR transcription, followed by speech captioning and timbre instruction generation. Phase 2 augments the corpus by training a speech-text embedding model for retrieval from internal TTS data and a fine-tuned caption model for scalable annotation.}
    \label{fig:pipeline}
\end{figure}

As shown in Fig.~\ref{fig:pipeline}, the data collection process includes two phases:

\paragraph{Phase 1: Collection and Annotation of Cinematic Data}

Motivated by the hypothesis that exposure to real-world acoustic variation produces more perceptually natural speech, we adopt a bottom-up approach in collecting audio from movies, TV dramas, and episodic series, as opposed to relying on scripted studio recordings. This type of content inherently offers diverse scenes, characters, and expressive speaking styles, providing a richer source of acoustic variation than controlled recording environments.

The data processing pipeline involves the following stages:
\begin{itemize}
    \item \textbf{Speaker Diarization}: We adopt DiariZen~\citep{han2025efficient,han2025fine,han2025leveraging} to identify and segment audio by different speakers.
    \item \textbf{Denoising and Quality Filtering}: We apply MossFormer2\_SE\_48K~\citep{zhao2023mossformer2} for noise reduction and use DNSMOS~\citep{reddy2021dnsmos} to filter out low-quality audio. 
    MossFormer2\_SE\_48K is particularly effective at extracting clean speech from challenging acoustic conditions, including significant background noise, music tracks, and speech-overlaid BGM, making it particularly suited for processing noisy and music-infused audio sources.
    
    To ensure the quality of the audio data, we set a DNSMOS threshold of $\geq$ 3.0 for quality filtering. Cinematic data often contains substantial background noise, and without denoising, only around 5\% of the samples meet the DNSMOS 3.0 threshold. After applying MossFormer2\_SE\_48K for denoising, the retention rate increases to approximately 45\%–50\%. Although this denoising process may result in minor quality loss (e.g., occasional breath noises or slight loss of high-frequency details), our priority is to maximize sample retention and preserve the diversity of speech, particularly given the scarcity of high-quality cinematic data.
    \item \textbf{Single Speaker Filtering}: Given the potential inaccuracies in diarization, especially in distinguishing speakers with similar voices (e.g., speakers of the same gender), we leverage MOSS-Transcribe-Diarize~\citep{ai2026mosstranscribediarizetechnical}, which excels at producing accurate per-speaker transcriptions in multi-speaker conversations, to identify and retain only single-speaker segments. This ensures that the model does not learn to generate inconsistent voice tones during speech synthesis.
    \item \textbf{ASR Transcription}: Through multiple iterations of ASR models, we eventually select Qwen3-Omni-30B-A3B-Instruct~\citep{Qwen3-Omni} for transcription. This model performs well in recognizing numbers, symbols, and standard punctuation, which is essential for capturing the emotional nuances in speech. We also apply simple rule-based filtering to remove empty and duplicate entries, which are usually caused by pure music or noise, leading to model collapse. Given that the corpus primarily consists of Chinese and English audio, we use Whisper-large-v3~\citep{radford2022whisper} to filter out non-Chinese and non-English languages. This decision is driven by the need for the model to effectively learn from a focused dataset, as introducing additional languages could compromise the quality and accuracy of transcriptions in the current scope. While our focus is on Chinese and English, we are committed to expanding this capability in the future to support low-resource languages, aiming to improve inclusivity and model robustness in TTS models.
\end{itemize}

The resulting dataset consists of approximately 5,000 hours of audio, which we then annotate using Gemini-2.5 Pro~\citep{comanici2025gemini} to generate detailed captions. These captions are converted into natural language timbre instructions using Qwen3-32B-A3B-Instruct~\citep{qwen3technicalreport}.

\paragraph{Phase 2: Data Augmentation}

In Phase 2, we scale up the dataset through two complementary strategies:
\begin{itemize}
    \item \textbf{Fine-Tuning a Speech Caption Model}: We fine-tune Qwen3-Omni-30B-A3B-Thinking~\citep{Qwen3-Omni} on the Phase 1 data to train a dedicated speech caption model, which will be used for automatic annotation of large-scale audio in the next steps.
    \item \textbf{Style-Guided Audio Mining}: We also train a speech-text alignment embedding model ~\citep{fanspeech}, which maps style instructions and speech audio into a shared embedding space, enabling text-based retrieval of audio clips that match a given style description. Because much of our internal TTS base training corpus consists of relatively neutral speech (e.g., audiobook narration), directly sampling from it would yield predominantly flat, expressively homogeneous data. We further leverage GPT-5~\citep{singh2025openai} to generate a diverse set of \emph{stylistically rich} text-based timbre instructions. 
    By using these style-oriented instructions as queries against the shared embedding space, we can surface recordings that exhibit salient expressive characteristics from within this otherwise neutral collection. For each instruction, we retrieve the top 50 matches ranked by cosine similarity and immediately remove them from the candidate pool before processing the next instruction, ensuring that no audio clip is selected more than once across different queries. This targeted mining adds approximately 10,000 hours of stylistically diverse audio data that complements the cinematic sources in Phase~1.
\end{itemize}

The newly matched audio is processed following the same steps as Phase 1, except that for cost efficiency, we replace the captioning model with the internal, custom-trained caption model developed during the fine-tuning process.
To further supplement coverage, we incorporate additional data from other internal sources, mainly crowdsourced dubbing recordings, annotated using the same custom-trained caption model, yielding the final dataset of approximately 25,000 hours.

\paragraph{Dataset Statistics}

\begin{figure}[t]
    \centering
    \includegraphics[width=1.0\textwidth]{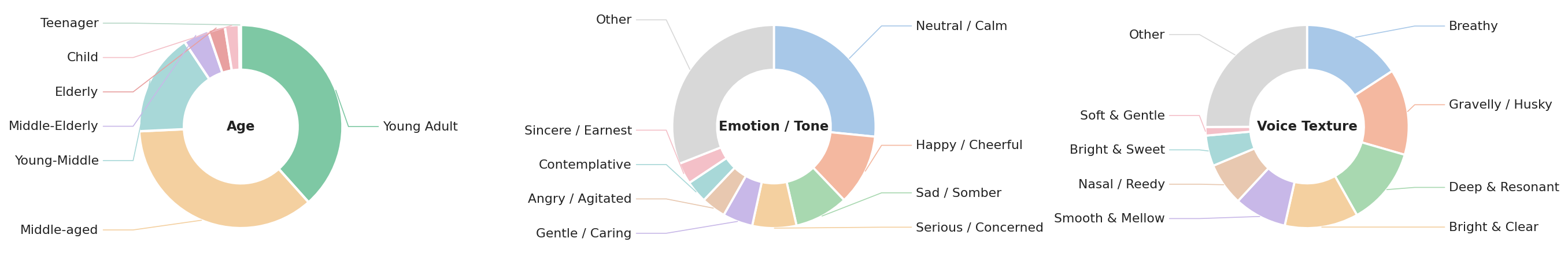}
    \caption{A snapshot of the training corpus profiled along three perceptual dimensions based on the caption results. The distributions reveal broad, naturalistic coverage of everyday speaking styles.}
    \label{fig:sunburst}
\end{figure}

The final dataset comprises approximately 25,000 hours of unique audio across Chinese (18,025 hours) and English (7,047 hours).
To better understand the composition of our corpus, we uniformly sample 150K utterances from each language and profile them along three perceptual dimensions: speaker age, emotion/tone, and voice texture, as illustrated in Fig.~\ref{fig:sunburst}. The resulting slices confirm that the data captures a naturally diverse range of everyday speaking characteristics—skewing toward young-adult and middle-aged speakers, predominantly neutral-to-positive tones, and a broad palette of voice textures.


\subsection{Training Strategy}

MOSS-VoiceGenerator starts from the Qwen3~\citep{qwen3technicalreport} checkpoint weights, and is trained end-to-end on our curated instruction-text-speech dataset. The model receives a structured input composed of a natural language timbre instruction and a target transcript, concatenated via a fixed chat template and encoded by the LLM's native text tokenizer. The training objective is standard next-token prediction loss over the codec token sequence, conditioned on the text input. All model parameters are updated during training; we do not apply parameter-efficient methods such as LoRA.


\paragraph{Model Scale}
We compare the 1.7B and 8B backbone sizes under the same training recipe. The 1.7B model achieves comparable instruction-following quality to the 8B model, while demonstrating better generation diversity. We suspect that the data volume might not have been sufficient, and thus, we attempted to mix in approximately 10,000 hours of TTS-base data (without instructions, pure text-speech pairs). However, we observed no significant gain in either objective benchmarks or human evaluation. As a result, we decided to adopt the 1.7B model, trained exclusively on instruction data, as the final released configuration.

\paragraph{English Prosody Augmentation}
Due to the limited scale of the English subset ($\sim$7,047 hours vs.\ $\sim$18,025 hours for Chinese), early checkpoints exhibited unstable English prosody characterized by frequent unnatural pauses. To address this, we apply instruction rewriting to all English samples: for each audio clip, we generate two semantically equivalent but lexically distinct instruction variants, thereby doubling the effective English training signal without requiring additional audio collection. This augmentation substantially improves English prosody stability in the final model.

\section{Evaluation}

\subsection{Objective Evaluation}

We evaluate MOSS-VoiceGenerator on InstructTTSEval~\citep{huang2025instructttseval}, a public benchmark designed to assess TTS models' ability to follow complex natural-language style instructions.
InstructTTSEval comprises 6{,}000 test cases (3 tasks $\times$ 2 languages $\times$ 1{,}000 samples) drawn from movies, TV dramas, and variety shows, each paired with a reference audio clip.
The benchmark defines three evaluation tasks of increasing abstraction:
\begin{itemize}
    \item \textbf{APS (Acoustic-Parameter Specification):} Instructions explicitly specify all 12 acoustic attributes (pitch, speed, emotion, gender, age, clarity, fluency, accent, texture, tone, volume, personality), testing direct fine-grained control.
    \item \textbf{DSD (Descriptive-Style Directive):} APS instructions are rewritten by an LLM into free-form natural language descriptions with random attribute omissions, testing generalization to incomplete and unstructured inputs.
    \item \textbf{RP (Role-Play):} Instructions provide abstract role and scenario descriptions (e.g., ``a nervous interview applicant''), requiring the model to infer appropriate vocal characteristics from context.
\end{itemize}

To prevent test-set contamination, we ensure that our training corpus does not overlap with the InstructTTSEval evaluation set. Additionally, we perform fuzzy matching on each transcript in the InstructTTSEval set to ensure that no test data from InstructTTSEval appears in our training data.

We compare against gemini-2.5-pro-preview-tts~(hereafter referred to as Gemini-TTS-Pro), GPT-4o-mini-TTS, VoxInstruct, MIMO-Audio-7B-Instruct, VoiceSculptor, and Qwen3-TTS-VD. Note that Gemini-TTS-Pro and GPT-4o-mini-TTS only support fixed voice editing rather than free-form instruction-driven voice design, and VoxInstruct requires reference audio.  Results for Gemini-TTS-Pro, GPT-4o-mini-TTS, and VoxInstruct are taken directly from the InstructTTSEval benchmark paper, while results for VoiceSculptor, MIMO-Audio-7B-Instruct, and Qwen3-TTS-VD are sourced from their respective technical reports. As shown in Tab.~\ref{tab:performance-en-zh}, MOSS-VoiceGenerator demonstrates competitive performance within the open-source landscape.

\begin{table}[tb]
    \centering
    \caption{Instruction-following accuracy (\%) on InstructTTSEval.}
    \begin{tabular}{l|ccc|ccc}
        \toprule
        \multirow{2}{*}{\textbf{Model}} &
        \multicolumn{3}{c|}{\textbf{InstructTTSEval-EN}} &
        \multicolumn{3}{c}{\textbf{InstructTTSEval-ZH}} \\
        & \textbf{APS} & \textbf{DSD} & \textbf{RP}
        & \textbf{APS} & \textbf{DSD} & \textbf{RP} \\
        \midrule
        \rowcolor{gray!15}
        Gemini-TTS-Pro & 87.6 & 86.0 & 67.2 & 89.0 & 90.1 & 75.5 \\
        \rowcolor{gray!15}
        GPT-4o-mini-TTS & 76.4 & 74.3 & 54.8 & 54.9 & 52.3 & 46.0 \\
        \rowcolor{gray!15}
        VoxInstruct & 54.9 & 57.0 & 39.3 & 47.5 & 52.3 & 42.6 \\
        VoiceSculptor-VD & - & - & - & 75.7 & 64.7 & 61.5 \\
        MIMO-Audio-7B-Instruct & 80.6 & 77.6 & 59.5 & 75.7 & 74.3 & 61.5 \\
        Qwen3-TTS-VD & 78.4 & 78.8 & 72.0 & 84.3 & 82.9 & 77.4 \\
        \midrule
        \rowcolor{MossCyan!15}
        \textbf{MOSS-VoiceGenerator} & 68.2 & 82.0 & 68.7 & 78.0 & 80.0 & 74.0 \\
        \bottomrule
    \end{tabular}
    \label{tab:performance-en-zh}
\end{table}

\subsection{Subjective Evaluation}
For subjective evaluation, we conduct a pairwise preference study on the same 100 instruction–speech pairs (50 Chinese, 50 English) drawn from an internal test set distinct from InstructTTSEval, covering diverse styles and roles. For each pair, human listeners are presented with the outputs of MOSS-VoiceGenerator and a baseline model generated from the same instruction, with audio presentation order randomized; listeners are blind to which model produced each output. They indicate their preference on three dimensions: Overall Performance, Instruction Following, and Naturalness. These dimensions are selected for their relevance to real-world use cases and are critical for evaluating the quality of the generated voices:

\begin{itemize}
\item \textbf{Overall Performance}: \emph{Given two audio outputs generated from the same instruction, which one would you ultimately prefer?}  This dimension reflects the holistic quality of the audio, encompassing factors such as human-likeness, emotional expressiveness, audio fidelity, naturalness of prosody, clarity of articulation, speaker consistency, and overall listening experience.
\item \textbf{Instruction Following}: \emph{Which output better follows the instruction?} Instructions may specify attributes such as gender, age, voice characteristics, emotion, accent, etc. For instructions involving specific characters or scenarios, this dimension also considers whether the voice matches your subjective expectation of that character or scene.
\item \textbf{Naturalness}: \emph{Which output sounds more like real human speech?} This dimension evaluates how closely the generated audio resembles natural human speech in terms of rhythm, pacing, intonation, and conversational quality.
\end{itemize}

Each item is annotated by three independent crowdsourced workers (100–300 items per worker, 1 RMB per item); the final label is determined by majority vote. For each pair, each annotator scores only one of the dimensions to ensure focus and minimize bias. This approach allows for a thorough and unbiased evaluation of each dimension independently.

We compare against Qwen3-TTS-VD, MiniMax Voice Design, and MiMo-Audio-7B-Instruct—models that provide publicly accessible voice design interfaces, suitable for pairwise listening studies. We exclude Gemini-TTS-Pro and GPT-4o-mini-TTS from the preference study, as these models offer preset voices rather than free-form voice design, which makes a direct comparison invalid. For Qwen3-TTS-VD and MiniMax, we use their respective APIs, with all models set to their default configurations, aside from the input voice descriptions and text. Similarly, MIMO-Audio-7B-Instruct is evaluated using the default settings as specified in its official tutorial. All models are evaluated with a single inference run.

\begin{figure}[tbp]
    \centering
    \includegraphics[width=1.0\textwidth]{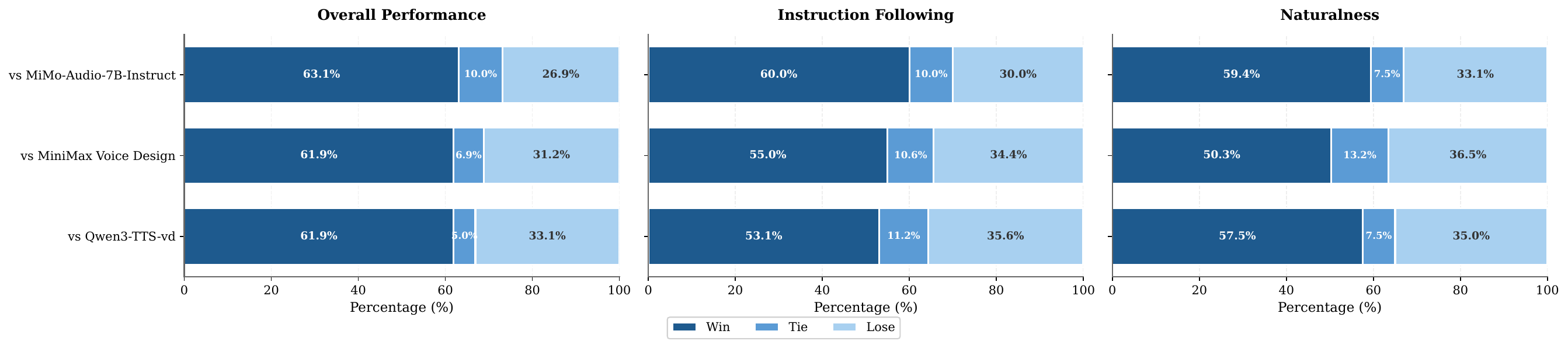}
    \caption{Pairwise preference results (Win / Tie / Lose) of MOSS-VoiceGenerator against three baselines across three evaluation dimensions. Each bar reports the percentage of comparisons won, tied, or lost. MOSS-VoiceGenerator consistently wins on all three dimensions against all baselines.}
    \label{fig:subjective}
\end{figure}

As shown in Fig.~\ref{fig:subjective}, MOSS-VoiceGenerator outperforms all three baselines across the three evaluation dimensions.
In our analysis, MOSS-VoiceGenerator excels at generating everyday conversational voices that exhibit natural pauses, hesitations, and rhythmic variations, clearly distinguishing them from the steady, broadcast-style delivery typical of newsreader speech. Furthermore, the model convincingly reproduces a wide range of authentic vocal traits—from the voices of the elderly and children to laughter, sobbing, awkwardness, and anger—delivering a level of realism that feels genuinely spontaneous rather than studio-recorded. These capabilities make MOSS-VoiceGenerator particularly well-suited for applications that demand expressive and immersive speech, such as audiobook narration, film dubbing, and emotionally nuanced voice assistants.

\section{Conclusions}
We presented \textbf{MOSS-VoiceGenerator}, an open-source instruction-driven TTS model that synthesizes realistic, expressive speech directly from natural language descriptions, without relying on any reference audio.
To train this model, we carefully source speech data from TV dramas and films, and construct a large-scale in-the-wild corpus with fine-grained annotation, capturing a remarkably diverse range of speaker identities, emotions, and acoustic conditions.
Benefiting from this expressive and diverse training corpus, MOSS-VoiceGenerator demonstrates strong voice design capabilities, and excels in subjective evalutaion, including overall quality, instruction-following accuracy, and speech naturalness.
We fully open-source our model, data pipeline, and evaluation toolkit, and hope this work provides the community with a practical and extensible foundation for advancing controllable, expressive TTS research.

\section{Limitations}

MOSS-VoiceGenerator has several limitations.
First, the language coverage is currently limited to Chinese and English. While the model performs well on these languages, extension to low-resource languages remains a focus for future work.
Second, while the English training corpus is sizable, it is still substantially smaller than the Chinese corpus. Despite using instruction rewriting augmentation, this discrepancy can sometimes result in weaker prosody in certain English-speaking styles, highlighting the need for further corpus expansion.
Third, our denoising-before-filtering pipeline, while increasing data retention, may introduce mild artifacts, such as residual breath noise or high-frequency smoothing. These artifacts can degrade performance in certain cases, though disentangling their perceptual impact from model capacity effects remains challenging.
Finally, the model's output can occasionally lack stability. We plan to enhance robustness in future versions to ensure more consistent and reliable generation across a wider range of inputs.

\section*{Contributors}

\noindent\textbf{Contributors}: \\
Kexin Huang$^{*}$, Liwei Fan, Botian Jiang, Yaozhou Jiang, Qian Tu, Jie Zhu, Yuqian Zhang, Yiwei Zhao, Chenchen Yang, Zhaoye Fei, Shimin Li, Xiaogui Yang, Qinyuan Cheng

\vspace{0.5em}

\noindent\textbf{Advisors}: \\
Xipeng Qiu$^{\S}$

\vspace{1em}

\noindent\textbf{Affiliations}: \\
Shanghai Innovation Institute\\
MOSI Intelligence\\
Fudan University\\

{\let\thefootnote\relax\footnotetext{$^*$Project lead: \url{kxhuang24@m.fudan.edu.cn}.\  $^\S$Corresponding author: \url{xpqiu@fudan.edu.cn}.}}

\clearpage
\bibliographystyle{unsrtnat}
\bibliography{main}

\end{document}